\documentclass[10pt, a4paper]{article}
\pdfoutput=1
\usepackage[utf8]{inputenc}
\usepackage[T1]{fontenc}
\usepackage{lmodern, textcomp}
\usepackage[british]{babel}
\usepackage{csquotes}

\usepackage[heightrounded, top=4cm, bottom=4cm, left=3.5cm, right=3.5cm]{geometry}

\usepackage{eurosym, xspace}
\usepackage[nottoc]{tocbibind}
\usepackage{graphicx, subfig, float}
\usepackage[multiple]{footmisc}
\usepackage{authblk}

\usepackage[usenames, dvipsnames]{xcolor}

\usepackage[backend=bibtex8, citestyle=numeric-comp, maxbibnames=99,
			sorting=none, sortcites=true, giveninits=true]{biblatex}
\input{arxiv.def}

\usepackage{amsmath, amsfonts, amssymb}
\usepackage{cancel}

\usepackage{maths_min}

\usepackage[pdftex, breaklinks=true, linktocpage,
	colorlinks=true, urlcolor=blue, linkcolor=blue, citecolor=red,
]{hyperref}
\usepackage[all]{hypcap}

\graphicspath{{images/}}

\DeclareUnicodeCharacter{00A0}{~}
\DeclareUnicodeCharacter{202F}{~}

\newcommand{\email}[1]{\thanks{\href{mailto:#1}{\nolinkurl{#1}}}}

\usepackage[sort&compress, english]{cleveref}

\hypersetup{
	pdftitle={Universality of tunnelling particles in Hawking radiation},
}

\title{Universality of tunnelling particles in Hawking radiation}

\author[1]{Harold Erbin\email{harold.erbin@physik.lmu.de}}
\affil[1]{Arnold Sommerfeld Center for Theoretical Physics, Ludwig--Maximilians--Universität München, Theresienstraße 37, 80333 München, Germany}

\author[2]{Vincent Lahoche\email{vincent.lahoche@labri.fr}}
\affil[2]{La\textsc{bri}, Université de Bordeaux, \textsc{Umr} 5800, 33405 Talence, France}

\begin{document}

\maketitle

\begin{abstract} 
The complex path (or Hamilton--Jacobi) approach to Hawking radiation corresponds to the intuitive picture of particles tunnelling through the horizon and forming a thermal radiation.
This method computes the tunnelling rate of a given particle from its equation of motion and equates it to the Boltzmann distribution of the radiation from which the Hawking temperature is identified.
In agreement with the original derivation by Hawking and the other approaches, it has been checked case by case that the temperature is indeed universal for a number of backgrounds and the tunnelling of particles from spins $0$ to $1$ mostly, spins $3/2$ and $2$ in some instances.
In this letter, we give a general proof that the temperature is indeed equal for all (massless and massive) particles with spins from $0$ to $2$ on an arbitrary background (limited to be Einstein for spin greater than $1$) in any number of dimensions.
Moreover, we propose a general argument to extend this result to any spin greater than $2$.
\end{abstract}

\newpage

\pdfbookmark[1]{\contentsname}{toc}
\tableofcontents
\bigskip
\hrule

\section{Introduction}

In his seminal paper~\cite{Hawking:1975:ParticleCreationBlack}, Hawking proved that black holes emit a thermal radiation at a temperature $T$ due to quantum mechanical effects.
More generally, thermal radiation is more generally associated to all horizons, including the ones of an accelerated observer (Unruh effect) or in cosmologies (cosmological horizons, e.g.\ in FRW universe or dS space).
In the following years, several other methods have been designed to compute such thermal effects and established the as major predictions of quantum field theories on curved spaces.
It is one of the rare instances where hints of a quantum gravity theory can be found, and as such, it is of primordial importance to understand it precisely.

The intuitive picture of this radiation is the following: pairs of virtual particles created near a black hole horizon through vacuum fluctuations become real once one of them cross the horizon while the other extracts energy from the black hole.
Two approaches realise this specific idea of tunnelling: the complex path (or Hamilton--Jacobi) method due to Shankaranarayanan--Srinivasan--Padmanabhan~\cite{Srinivasan:1999:ParticleProductionComplex, Shankaranarayanan:2001:MethodComplexPaths, Shankaranarayanan:2002:HawkingRadiationDifferent} (see also~\cite{Angheben:2005:HawkingRadiationTunneling}), and the null geodesic method (or Parikh--Wilczek) method~\cite{Parikh:2000:HawkingRadiationTunneling} (see~\cite{Vanzo:2011:TunnellingMethodsHawkings} for a review).
Both methods are not restricted to black hole radiation but can also be applied to any black hole with a thermal horizon or any other background which can have a thermal horizon, such as the Rindler or de Sitter spaces.
Moreover, they can also be used to define the Hawking temperature in situations where the traditional methods are not defined~\cite{DiCriscienzo:2010:HamiltonJacobiMethodGravitation}.

The complex path formalism computes the tunnelling rate of a particle of a given spin $s$ by solving its equations of motion in the black hole background through a WKB approximation, and then equates this rates to the probability given by the Boltzmann distribution at temperature $T$.
From the other methods, it is clear that the Hawking temperature $T$ is universal, i.e.\ that it is a property of the black hole and not of the tunnelling particle.\footnotemark{}
\footnotetext{%
    In particular, because most of these other approaches don't require to specify the type of the tunnelling particle.
}%
The main drawback of the complex path method is to hide this fact, in that the computations depend strongly on the tunnelling particle under consideration (and in particular on its spin, since the starting point is the equation of motion of the associated field).
Nonetheless it has been checked explicitly case by case for many backgrounds that the tunnelling of particles with different spins (mostly $s = 0, 1/2, 1$, but also $s = 3/2, 2$ in some instances) always yields the same temperature, see~\cite{Kerner:2006:TunnellingTemperatureTaubNUT, Kerner:2008:FermionsTunnellingBlack, Kerner:2008:ChargedFermionsTunnelling, Majhi:2010:HawkingRadiationDue, Yang:2014:HawkingRadiationBlack, Siahaan:2015:HigherSpinsTunneling, Sakalli:2016:BlackHoleRadiation} for a selected sample and references therein for more details.
As a consistency check, it would be desirable to establish the universality of the Hawking temperature in the complex path formalism in full generality.

The goal of this letter is to prove this result for neutral massless and massive particles with spin ranging from $0$ to $2$ for a generic background (restricted to be Einstein for $s = 3/2, 2$) in any dimension, completing results obtained earlier in~\cite{Yang:2014:HawkingRadiationBlack}.
This is achieved in two steps.
First, the equation of motion for a spin $s \le 2$ is rewritten into a second-order equation together with constraints following standard methods (this is the usual starting point to the analysis of the degrees of freedom)~\cite{Parker:2009:QuantumFieldTheory, Freedman:2012:Supergravity}.
Second, we show that, in the WKB approximation, this second-order equation reduces to the Hamilton--Jacobi equation of a scalar field (or, said differently, that the eikonal limit of field equations is universal), a fact which is certainly to be expected.
These computations hold for any background spacetime, and thus more particularly for the ones which have an horizon and for which one wishes to apply the complex path method.\footnotemark{}
\footnotetext{%
	In order to make the paper (almost) self-contained, we provide a short summary of the complex path method, including a description of the context and of the main steps of the derivation starting from the scalar Hamilton--Jacobi equation.
	Nonetheless, we do not provide any specific examples to avoid wandering too far from the main topic of this paper.
	We refer the reader interested to the method itself to the vast literature, and, more specifically, to the excellent review~\cite{Vanzo:2011:TunnellingMethodsHawkings}.
}%
We then give a general argument to extend this argument to massive particles of any spin $s > 2$.
Moreover, we stress that our proof is fully covariant, in contrast with the former computations which were not explicitly covariant since the fields and the background metric were written in components.

The limitation on the background for spins $s > 1$ and the need of non-minimal coupling are related to the well-known difficulty propagating higher-spin particles on a curved background~\cite{Buchdahl:1958:CompatibilityRelativisticWave, Cohen:1967:CovariantWaveEquations, Aragone:1971:ConstraintsGravitationallyCoupled} and in itself is not related to Hawking radiation.

An interesting question would be to analyse the subleading quantum corrections and the deviation from thermality due to the backreaction of the radiation and to see how they differ for the different types of particles (the greybody factor is definitely not universal).
The generalization of our argument to background with gauge fields under which the particles are charged is expected to be straightforward, even if one can expect difficulties already for $s = 1$ due to inconsistencies in the coupling of spin $s \ge 1$ to electromagnetic fields~\cite{Fierz:1939:RelativisticWaveEquations, Velo:1969:NoncausalityOtherDefects}.

In \cref{sec:complex-path} we review the complex path method for a scalar field and we show in \cref{sec:higher-spins} how the higher-spin cases reduce to this case.

\section{Complex path method for the scalar field}
\label{sec:complex-path}

In this section, we sketch the essential steps of the derivation of the Hawking temperature from the field equation of a scalar field in the complex path formalism.
In particular, we do not focus on a specific background to avoid overwhelming the reader with details not relevant to the derivation of the main result of this paper in \Cref{sec:higher-spins}.
The reader is referred to the literature~\cite{Srinivasan:1999:ParticleProductionComplex, Shankaranarayanan:2001:MethodComplexPaths, Shankaranarayanan:2002:HawkingRadiationDifferent, Angheben:2005:HawkingRadiationTunneling, Kerner:2006:TunnellingTemperatureTaubNUT, Vanzo:2011:TunnellingMethodsHawkings} (and references therein) for complete explanations and specific examples.

One considers a background (case of interests being black holes, the Rindler space, etc.) in $d$ dimensions described by a fixed metric $g_{\mu\nu}$.
For definiteness, the background metric is taken to be a solution of the Einstein equations with a cosmological constant
\begin{equation}
	\label{eom:einstein-equation}
	R_{\mu\nu} - \frac{1}{2}\, g_{\mu\nu} R + \Lambda g_{\mu\nu} = 0\,, 
\end{equation} 
where $\Lambda$ is the cosmological constant, $R_{\mu\nu} = \tensor{R}{_{\rho\mu}^\rho_\nu}$ is the Ricci tensor obtained by contracting the Riemann tensor, and $R = g^{\mu\nu} R_{\mu\nu}$ is the Ricci scalar.
As we will see, this restriction to Einstein spacetimes for deriving the Hawking temperature only concerns spins higher than $1$.
The Laplacian on this background is defined by
\begin{equation}
	\lap = g^{\mu\nu} \grad_\mu \grad_\nu
\end{equation} 
where $\grad_\mu$ is the covariant derivative with the Levi--Civita connection.
The tunnelling rate $\Gamma$ for a particle is given by
\begin{equation}
	\label{eq:tunnelling-rate}
	\Gamma
		= \frac{P_{\text{out}}}{P_{\text{in}}}
		= \frac{\abs{\phi_{\text{out}}}^2}{\abs{\phi_{\text{in}}}^2}
\end{equation} 
where $P_{\text{in}}$ ($P_{\text{out}}$) is the tunnelling probability for an ingoing (outgoing) particle and $\phi_{\text{in}}$ ($\phi_{\text{out}}$) is the associated solution to the equation of motion.
Assuming that the radiation is thermal\footnotemark{} this rate can be equated to the Boltzmann distribution through the detailed balance
\footnotetext{%
	This hypothesis is not strictly correct due to backreaction of the radiation on the geometry~\cite{Parikh:2000:HawkingRadiationTunneling}, but we will ignore this effect for our purpose.
}%
\begin{equation}
	\Gamma = \e^{- E_{\text{tot}} / T}
\end{equation} 
where $E_{\text{tot}}$ is the total energy (including kinetic, rotational, electromagnetic, etc.) carried by the particle tunnelling, and measured by a freely falling observer in the vicinity of the external horizon.

From this point, we consider a free (massive or massless) spin $0$ scalar field $\phi$.
The equation of motion for a scalar field in a curved background with non-minimal coupling
\begin{equation}
	\label{eom:phi}
	\left(- \lap + \frac{m^2}{\hbar^2} + \xi R \right) \phi = 0
\end{equation} 
where $m^2$ can be zero.
This equation can be solved at leading order in $\hbar$ through the WKB approximation
\begin{equation}
	\label{eq:phi-wkb}
	\phi(x) = \phi_0\, \e^{i S(x) / \hbar}\,,
\end{equation} 
where $\phi_0$ is a constant wave function.
Inserting this ansatz into \eqref{eom:phi} provides, at leading order in $\hbar$, the Hamilton--Jacobi equation on curved space
\begin{equation}
	\label{eq:hamilton-jacobi-scalar}
	g^{\mu\nu} \pd_\mu S \pd_\nu S + m^2 = 0\,,
\end{equation}
and allows to identify $S$ with the classical action and one can note that the non-minimal coupling term is subleading (such terms are also present for higher spins and will not contribute at leading order).

In terms of $S_{\text{in}}$ and $S_{\text{out}}$ the tunnelling rate \eqref{eq:tunnelling-rate} reads
\begin{equation}
	\Gamma
		= \frac{\abs{\phi_{\text{out}}}^2}{\abs{\phi_{\text{in}}}^2}
		= \e^{- 2 (\Im S_{\text{out}} - \Im S_{\text{in}}) / \hbar} \,.
\end{equation} 
The functions $S_{\text{in}}$ and $S_{\text{out}}$ can be solved quite generically with the following ansatz~\cite{Angheben:2005:HawkingRadiationTunneling, Kerner:2006:TunnellingTemperatureTaubNUT, Kerner:2008:ChargedFermionsTunnelling}
\begin{equation}
	S_{\text{out}} = - E t + W(r_0) + F(x^i) + K, \qquad
	S_{\text{in}} = - E t - W(r_0) + F(x^i) + K\,,
\end{equation} 
where $t$ is the time, $r_0$ the radial location of the horizon and $x^i$ denotes any other coordinate; $K$ is a complex constant, $W$ is complex and $F$ is real.
One needs to ensure that the ingoing probability is one in the classical limit because the horizon necessarily absorbs the particle.
This condition manifests itself differently depending on the choice of coordinates.\footnotemark{}
\footnotetext{%
	This point involves different subtleties and making a precise statement is very coordinate-dependent.
	We refer the reader to the literature for more details~\cite{Vagenas:2001:ComplexPathsCovariance, Angheben:2005:HawkingRadiationTunneling, Akhmedov:2006:HawkingTemperatureTunneling, Mitra:2007:HawkingTemperatureTunnelling, Akhmedov:2008:SubtletiesQuasiclassicalCalculation, Stotyn:2009:ObserverDependentHorizon}.
}%
It may occur that the inverse of the radial velocity has no pole for an ingoing classical particle, implying that this imaginary part vanishes.
If this is not the case then one needs to find the relation between the ingoing and outgoing actions such that this condition holds.
Both situations amount to setting $\Im K = \Im W(r_0)$ and one finally obtains the tunnelling rate
\begin{equation}
	\Gamma
		= \e^{- 4 \Im W(r_0) / \hbar}\,,
\end{equation} 
which yields the temperature
\begin{equation}
	\label{eq:temperature}
	T = \frac{\hbar E_{\text{tot}}}{4 \Im W(r_0)}
\end{equation} 
by equating with \eqref{eq:tunnelling-rate}.
In order to make contact with the well-known formula of the Hawking radiation, one can show (see for example~\cite{Kerner:2008:FermionsTunnellingBlack, Majhi:2010:HawkingRadiationDue, Vanzo:2011:TunnellingMethodsHawkings}) for general rotating black holes (including the Schwarzschild back hole as a limiting case) that the expression for $W(r_0)$ is proportional to the surface gravity $\kappa$:
\begin{equation}
	\Im W(r_0) = \frac{\pi E_{\text{tot}}}{2 \kappa}\,,
\end{equation} 
and the final result agrees with the well know Hawking temperature formula~\cite{Hawking:1975:ParticleCreationBlack}
\begin{equation}
	T = \frac{\hbar \kappa}{2\pi}\,.
\end{equation} 
The reason is that $W(r_0)$ is defined by an integral over $r$ with a pole at the horizon due to the presence of the metric components in the denominator: evaluating the integral yields a residue (imaginary) proportional to the surface gravity.
In the case of Schwarzschild one finds 
\begin{equation}
	f(r) = 1 - \frac{2 M}{r}, \qquad
	r_0 = 2 M
	\quad \Longrightarrow \quad
	\kappa = \frac{f'(r_0)}{2} = \frac{1}{4 M}, \qquad
	T = \frac{\hbar}{8\pi M}\,.
\end{equation}

\section{Tunnelling of higher-spins}
\label{sec:higher-spins}

In this section -- which contains our new results -- we show that the equations of motion for higher-spin particles reduce to the Hamilton--Jacobi equation \eqref{eq:hamilton-jacobi-scalar} of a scalar field in the leading order of the WKB approximation.
This is sufficient to establish that the temperature will be given by \eqref{eq:temperature} and thus that it is identical for all spins.\footnotemark{}
\footnotetext{%
	For this, it is important that the evaluation of the action in \eqref{eq:temperature} depends only on the properties of the background and not on the type of the particle.
}%
We stress that the computations of this section are valid for any QFT on a curved spacetime~\cite{Parker:2009:QuantumFieldTheory, Freedman:2012:Supergravity} (and thus not only the ones for which a Hawking temperature can be defined) and independent of the previous section.

\paragraph{Spin $1/2$}

The equation of motion for a spin $1/2$ fermion $\psi$ is
\begin{equation}
	\label{eom:psi-direct}
	\left( \slashed\grad - \frac{m}{\hbar} \right) \psi = 0
\end{equation} 
where $\slashed\grad \equiv \gamma^\mu \grad_\mu$ and $\gamma^\mu$ are the Dirac matrices.
The multiplication of \eqref{eom:psi-direct} with $\slashed\grad$ gives the second-order partial derivative equation:
\begin{equation}
	\label{eom:psi-final}
	- \lap \psi + \frac{1}{4}\, R \psi + \frac{m^2}{\hbar^2}\, \psi = 0.
\end{equation} 
As for the scalar field, the WKB approximation for this equation can be investigated using the following ansatz
\begin{equation}
	\label{eom:psi-wkb}
	\psi(x) = \psi_0(x)\, \e^{i S(x) / \hbar}\,,
\end{equation} 
where $\psi_0$ is a position-dependent spinor.
Putting this ansatz in \eqref{eom:psi-final}, we deduce an equation for $S$, and keeping only the leading order terms in $\hbar$, it reduces to the the scalar Hamilton--Jacobi equation \eqref{eq:hamilton-jacobi-scalar}.
In particular, no derivative of $\psi_0$ appears because it would be subleading in $\hbar$.

The Hamilton--Jacobi equation \eqref{eq:hamilton-jacobi-scalar} can also be derived by plugging \eqref{eom:psi-wkb} directly inside the Dirac equation \eqref{eom:psi-direct} and squaring the equation
\begin{equation}
	\label{eq:hamilton-jacobi-1st}
	(i \slashed\pd S - m) \psi_0 = 0.
\end{equation} 
Here also there is no derivative of $\psi_0$ because it is subleading in $\hbar$.
Note that it is necessary to keep the spinor $\psi_0$ when writing the first-order equation because the operator is not diagonal (as it is in the Klein--Gordon equation).

Before moving to the other spins, it is useful to develop the last point and to make a comment about the solutions of the equations.
Since the Dirac equation \eqref{eom:psi-direct} is of first-order, it is stronger than the modified Klein--Gordon equation \eqref{eom:psi-final} and, as such, not all solutions of the latter are solutions of the former (for $s > 1/2$ the second-order equations will be accompanied with constraints).
However, the converse is true and it is this fact which is relevant here.
First, the Hamilton--Jacobi equation \eqref{eom:psi-wkb} can be used to determine $S$.
To get a solution of the original Dirac equation (to leading order in $\hbar$), one needs to check that \eqref{eq:hamilton-jacobi-1st} is satisfied.
This is achieved by inserting the solution for $S$ found from the Hamilton--Jacobi equation and by solving for $\psi_0$ (see~\cite[sec.~2.2.3]{Itzykson:2006:QuantumFieldTheory} for examples of this method).
Since $\psi_0$ is a general position-dependent spinor, a solution generically exists.
Since only $S$ is relevant to compute the tunnelling rate, we can safely ignore the computations of the constant amplitude $\psi_0$.
Nonetheless, for comprehensiveness, more precise conditions on the components are obtained in \Cref{sec:wave-conditions}.

While the same comment holds for half-integer spin particle, the argument is slightly different for integer spin particles; we will discuss it in the next section for $s = 1$.

\paragraph{Spin $1$}

The equation of motion for a massive vector field $A_\mu$ can be derived from the standard Proca Lagrangian, and may be written in the  form
\begin{subequations}
\begin{align}
	\label{eom:A-direct}
	0 = - \lap A_\mu + \grad_\nu \grad_\mu A^\nu + \frac{m^2}{\hbar^2}\, A_\mu.
\end{align}
\end{subequations}
Up to straightforward manipulations, this equation is equivalent to
\begin{equation}
	\label{eom:A-final}
	- \lap A_\mu + R_{\mu\nu} A^\nu + \frac{m^2}{\hbar^2}\, A_\mu = 0
\end{equation} 
together with the constraint
\begin{equation}
	\label{eq:grad-A-vanish}
	\grad_\mu A^\mu = 0
\end{equation}
which can be imposed at the dynamical level as a consequence of the equation of motion for $m^2 \neq 0$, or through a gauge transformation
\begin{equation}
	\delta A_\mu = \grad_\mu \alpha,
\end{equation} 
for vanishing mass, the scalar field $\alpha$ being the gauge parameter.
Remember that this constraint is necessary for ensuring that the correct degrees of freedom propagate (the spin $1$) while the extraneous ones are removed (the spin $0$ part).
In the leading order of the WKB approximation
\begin{equation}
	A_\mu(x) = A_{0\mu}(x) \, \e^{i S(x) / \hbar}
\end{equation} 
the equation \eqref{eom:A-final} corresponds to the scalar Hamilton--Jacobi equation \eqref{eq:hamilton-jacobi-scalar}.
Again, derivatives of $A_{0\mu}(x)$ do not arise because they are subleading.

As described in the previous section, one needs to ensure that solutions to \eqref{eom:A-final} are solutions to the original equation \eqref{eom:A-direct} (in the given approximation).
The WKB approximation of \eqref{eom:A-direct} reads
\begin{equation}
	A_{0\rho} \Big(
			g^{\mu\nu} \pd_\mu S \pd_\nu S + m^2
			\Big)
		- A_{0}^\sigma \pd_\sigma S \pd_\rho S
		= 0.
\end{equation} 
The first parenthesis vanishes as a consequence of the Hamilton--Jacobi equation, while the second term is zero due to the constraint \eqref{eq:grad-A-vanish}.
As a consequence, every solution of the Hamilton--Jacobi equation is also a solution of \eqref{eom:A-direct} (to the given approximation).

\bigskip

Note that for the two previous cases it was not necessary to use the fact that the background metric is a solution of the Einstein equation \eqref{eom:einstein-equation}.
Hence the universality of Hawking temperature for spin $s = 0, 1/2, 1$ is valid for any background, irrespective of the theory of gravity or the matter content under consideration, with the exception of gauge couplings.\footnotemark{}
\footnotetext{%
	Indeed the coupling to the gauge field in the covariant derivative comes with a factor $\hbar^{-1}$.
	On the other hand, couplings to other scalar and fermions fields can come only with positive powers of $\hbar$, implying that these terms do not contribute at the leading order of the WKB approximation.
}%
As noted in the introduction, the analysis of the second-order field equations for $s = 3/2$ and $s = 2$ shows that the background is restricted to be Einstein backgrounds; this stems from the well-known problem of propagating consistently fields with $s > 1$ on curved backgrounds.\footnotemark{}
\footnotetext{%
	However, one can expect the Hamilton--Jacobi equation to be identical for all particles since the spin has no effect in the WKB approximation.
	Indeed, in view of the local flatness of spacetime, one can use normal coordinates locally, and the field equations reduce to the one on Minkowski spaces.
	Then, one can extract the constraints on the field easily without having to use the Einstein equation \eqref{eom:einstein-equation}, and the Hamilton--Jacobi equation \eqref{eq:hamilton-jacobi-scalar} follows.
}%

\paragraph{Spin $3/2$}

The massive Rarita--Schwinger field is described by a (bi-)spinor-valued vector field $\psi_\mu$ whose equation of motion is:
\begin{equation}
	\label{eom:rarita}
	\gamma^{\mu\nu\rho} \grad_\nu \psi_\rho - \frac{m}{\hbar}\, \gamma^{\mu\nu} \psi_\nu = 0.
\end{equation} 
Some lengthy but simple manipulations~\cite{Amsel:2009:SupergravityBoundaryAdS} show that $\psi_\mu$ obeys the Dirac equation
\begin{equation}
	\label{eom:rarita-dirac}
	\left( \slashed\grad - \frac{m}{\hbar} \right) \psi_\mu = 0\,,
\end{equation} 
together with the condition
\begin{equation}
	\label{eq:constraint-slash-psi}
	\gamma^\mu \psi_\mu = 0\,.
\end{equation} 
Note that these conditions result from the equation of motion \eqref{eom:rarita} if
\begin{equation}
	m^2 \neq 0, m_0^2 \,, \qquad
	m_0^2 \equiv \frac{d - 2}{2 (d - 1)}\, \hbar^2 \Lambda\,,
\end{equation}
or from the gauge invariance under the following transformation otherwise:
\begin{equation}
	\delta \psi_\mu = \left(\grad_\mu - \frac{m_0}{(d - 2)}\, \gamma_\mu \right) \epsilon\,,
\end{equation} 
where $\epsilon$ is a spinor-valued gauge parameter.
As discussed for the spin $1$, the constraint ensures that only the spin $3/2$ part of the field propagates.
However, it can be imposed only if the background satisfies the Einstein equation \eqref{eom:einstein-equation} for the spin $3/2$.
Finally the equation \eqref{eom:rarita-dirac} can be multiplied with $\slashed\grad$ which leads to
\begin{equation}
	\label{eom:rarita-final}
	- \lap \psi_\rho
		+ \gamma^{\mu\nu} \tensor{R}{_{\mu\nu}^\sigma_\rho} \psi_\sigma
		+ \frac{R}{4}\, \psi_\rho
		+ \frac{m^2}{\hbar^2}\, \psi_\rho
		= 0\,,
\end{equation} 
and inserting the WKB ansatz
\begin{equation}
	\label{eq:rarita-wkb}
	\psi_\mu(x) = \psi_{0\mu}(x) \, \e^{i S(x) / \hbar}
\end{equation} 
inside the equation \eqref{eom:rarita-final} brings it to the form of \eqref{eq:hamilton-jacobi-scalar} at the leading order in $\hbar$.

Alternatively, it is possible to bypass the need of an Einstein background by considering the WKB approximation \eqref{eq:rarita-wkb} directly of the Dirac equation \eqref{eom:rarita-dirac}.
Then the equation reduces to the first-order equation \eqref{eq:hamilton-jacobi-1st} which leads immediately to the Hamilton--Jacobi equation \eqref{eq:hamilton-jacobi-scalar}.\footnotemark{}
\footnotetext{%
	In this case, the background must still be Einstein and the constraints have also to be imposed.
}%
This also shows that a solution to the Hamilton--Jacobi equation will be a solution of the original equation \eqref{eom:rarita} (in the WKB approximation), after solving for the constant vector-spinor $\psi_{0\mu}$.

\paragraph{Spin $2$}

The massive spin $2$ field is usually described by a symmetric tensor of rank $2$, $h_{\mu\nu}$, whose equation of motion may be written as ~\cite{Buchbinder:2000:EquationsMotionMassive}
\begin{multline}
	\label{eom:spin2}
	- \lap h_{\mu\nu}
		+ g_{\mu\nu} \lap h
		- \grad_\mu \grad_\nu h
		- g_{\mu\nu} \grad^\rho \grad^\sigma h_{\rho\sigma}
		+ \grad^\rho \grad_\mu h_{\nu\rho}
		+ \grad^\rho \grad_\nu h_{\mu\rho}
		- \frac{2\xi}{d}\, R\, h_{\mu\nu} \\
		- \frac{1 - 2\xi}{d}\, R\, h\, g_{\mu\nu}
		+ \frac{m^2}{\hbar^2}\, \big(h_{\mu\nu} - h g_{\mu\nu} \big)
		= 0
\end{multline} 
where $\xi$ is an arbitrary parameter parametrizing the non-minimal coupling (the latter is necessary in order to get the correct constraints on the propagating degrees of freedom below).
Then the equation \eqref{eom:spin2} can be simplified to
\begin{equation}
	- \lap h_{\mu\nu} - 2 \tensor{R}{^\rho_\mu^\sigma_\nu} h_{\rho\sigma} - \frac{2 (\xi - 1)}{d}\, R h_{\mu\nu} + \frac{m^2}{\hbar^2} h_{\mu\nu} = 0
\end{equation} 
together with the constraints
\begin{equation}
	\label{eq:spin2-constraints}
	h = 0, \qquad
	\grad^\mu h_{\mu\nu} = 0
\end{equation} 
if the background satisfies the Einstein equation \eqref{eom:einstein-equation}.
In the case where the condition
\begin{equation}
	m^2 \neq m_0^2 \equiv - \frac{4 \hbar^2 (1 - \xi)}{d - 2}\, \Lambda
\end{equation} 
holds, then the constraints \eqref{eq:spin2-constraints} result from the equation of motion \eqref{eom:spin2}~\cite{Buchbinder:2000:EquationsMotionMassive}.
Otherwise, if $m^2 = m_0^2$ then they can be imposed through a gauge transformation
\begin{equation}
	\label{eq:spin2-gauge-transformation}
	\delta h_{\mu\nu} = \grad_\mu \zeta_\nu + \grad_\nu \zeta_\mu\,.
\end{equation} 
Note that this includes the case of the graviton propagating on a curved space which corresponds to $m^2 = 0$ and $\xi = 1$~\cite{Deser:1984:GaugeInvarianceMasslessness}.\footnotemark{}
\footnotetext{%
	To our knowledge the gauge transformation \eqref{eq:spin2-gauge-transformation} has not been discussed elsewhere for generic $\xi$.
}%
In the WKB approximation
\begin{equation}
	h_{\mu\nu}(x) = h_{0\mu\nu}(x) \, \e^{i S(x) / \hbar}
\end{equation} 
the equation \eqref{eom:spin2} is again equivalent to \eqref{eq:hamilton-jacobi-scalar}.
Moreover, it is straightforward to check that this provides a solution to WKB approximation of the original equation \eqref{eom:spin2} by using the constraints \eqref{eq:spin2-constraints}.

\paragraph{Higher spins}

More generally one can consider a massive particle of arbitrary integer spin $s > 2$ (the case of half-integer is a straightforward extension) represented by a field $\phi_{\mu_1 \cdots \mu_s}$ symmetric in all indices for which the equation of motion is
\begin{equation}
	- \lap \phi_{\mu_1 \cdots \mu_s} + \tensor{f(R)}{_{\mu_1 \cdots \mu_s}^{\nu_1 \cdots \nu_s}}\, \phi_{\nu_1 \cdots \nu_s} + \frac{m^2}{\hbar^2}\, \phi_{\mu_1 \cdots \mu_s} = 0
\end{equation} 
after elimination of the auxiliary fields and imposing the constraints~\cite{Singh:1974:LagrangianFormulationArbitrary, Singh:1974:LagrangianFormulationArbitrary-1, Buchbinder:2005:BRSTApproachLagrangian}
\begin{equation}
	\grad^\mu \phi_{\mu \mu_2 \cdots \mu_s} = 0, \qquad
	g^{\mu \nu} \phi_{\mu \nu \mu_3 \cdots \mu_s} = 0,
\end{equation} 
where $f(R)$ is a function of the Riemann tensor and its contractions, arising both from anticommutation of covariant derivatives and from non-minimal coupling terms (which ensures causality and unitarity~\cite{Cucchieri:1995:TreeLevelUnitarityConstraints}).
Introducing the WKB ansatz
\begin{equation}
	\phi_{\mu_1 \cdots \mu_s}(x) = \phi_{0, \mu_1 \cdots \mu_s}(x) \, \e^{i S(x) / \hbar}
\end{equation} 
yields the Hamilton--Jacobi equation \eqref{eq:hamilton-jacobi-scalar}.
The reason is that curvature terms cannot have factors of $\hbar$ because they do not contain derivatives as the Laplacian or built-in factors as the mass term.
Any other term would be eliminated by the constraints (which are necessary for the theory to exist and be consistent).

\section*{Acknowledgments}

We are grateful to Eric Huguet and Vincent Rivasseau for reading the draft of the manuscript, and, more particularly, to Matt Visser for extensive comments.
H.E.\ acknowledges support from the \textsc{Cnrs} and the \textsc{Lptens} during the realization of this project.
The work of H.E.\ is conducted under a Carl Friedrich von Siemens Research Fellowship of the Alexander von Humboldt Foundation for postdoctoral researchers.

\appendix

\section{Conditions on spin-1/2 wave function}
\label{sec:wave-conditions}

In this section we derive conditions on the amplitude $\psi_0$ and envelope $S$ of the spin-$1/2$ WKB ansatz \eqref{eom:psi-wkb}
\begin{equation}
	\psi = \psi_0 \, \e^{i S / \hbar}
\end{equation} 
by studying the equation \eqref{eq:hamilton-jacobi-1st}
\begin{equation}
	(i \slashed\pd S - m) \psi_0 = 0.
\end{equation} 
We will follow the steps from~\cite{Kerner:2008:FermionsTunnellingBlack}, but our conditions are valid for all backgrounds since the expressions are given in the Lorentz frame.
The conventions follow~\cite[chap.~2]{Freedman:2012:Supergravity}.

Introduce a vielbein basis $e^a_\mu$, where $a = 0, \ldots, 3$ are the Lorentz frame indices, such that
\begin{equation}
	g_{\mu\nu} = \eta_{ab} e^a_\mu e^b_\nu.
\end{equation} 
The Dirac matrices in the chiral basis read
\begin{equation}
	\gamma^a =
	\begin{pmatrix}
		0 & \sigma^a \\
		\bar\sigma^a & 0
	\end{pmatrix},
	\qquad
	\sigma^a = (1, \sigma^i),
	\qquad
	\bar\sigma^a = (- 1, \sigma^i).
\end{equation} 
We recall the Pauli matrices $\sigma^i$
\begin{equation}
	\sigma^1 =
	\begin{pmatrix}
		0 & 1 \\
		1 & 0
	\end{pmatrix},
	\qquad
	\sigma^2 =
	\begin{pmatrix}
		0 & -i \\
		i & 0
	\end{pmatrix},
	\qquad
	\sigma^3 =
	\begin{pmatrix}
		1 & 0 \\
		0 & -1
	\end{pmatrix}.
\end{equation} 
One chooses to measure the spin in the $3$ direction.
The chiral up and down $2$-spinors $\xi_\pm$ which are eigenvectors of $\sigma^3$ read
\begin{equation}
	\xi_+ =
	\begin{pmatrix}
		1 \\ 0
	\end{pmatrix},
	\qquad
	\xi_- =
	\begin{pmatrix}
		0 \\ 1
	\end{pmatrix}
\end{equation} 
which leads to the ansatz
\begin{equation}
	\psi_0 =
	\begin{pmatrix}
		A_\pm \xi_\pm \\
		\pm i \, B_\pm \xi_\pm
	\end{pmatrix}
\end{equation} 
where $A_\pm$ and $B_\pm$ are constant numbers.

Inserting the ansatz in the first-order equation in the WKB approximation leads to
\begin{equation}
	\begin{pmatrix}
		- m & i \sigma^a \pd_a \\
		i \bar\sigma^a \pd_a & - m
	\end{pmatrix}
	\begin{pmatrix}
		A_\pm \xi_\pm \\
		\pm i \, B_\pm \xi_\pm
	\end{pmatrix}
		= 0
\end{equation} 
which splits in two equations
\begin{equation}
	(B_\pm \, \sigma^a \pd_a S \pm m A_\pm) \xi_\pm = 0,
	\qquad
	(A_\pm \, \bar\sigma^a \pd_a S \mp m B_\pm) \xi_\pm = 0.
\end{equation} 
To go further, one needs to write the equations in components, and we focus on the spin up case.
The first equation becomes
\begin{equation}
	B_+
		\begin{pmatrix}
			(\pd_0 + \pd_3) S & (\pd_1 - i \pd_2) S \\
			(\pd_1 + i \pd_2) S & (\pd_0 - \pd_3) S
		\end{pmatrix}
		\begin{pmatrix}
			1 \\ 0
		\end{pmatrix}
		+ m A_+
		\begin{pmatrix}
			1 \\ 0
		\end{pmatrix}
	= 0,
\end{equation} 
leading to the two equations
\begin{equation}
	B_+ (\pd_0 + \pd_3) S + m A_+ = 0,
	\qquad
	B_+ (\pd_1 + i \pd_2) S = 0.
\end{equation} 
The second equation gives
\begin{equation}
	A_+ (- \pd_0 + \pd_3) S - m B_+ = 0,
	\qquad
	A_+ (\pd_1 + i \pd_2) S = 0.
\end{equation} 

One first finds a constraint on $S$
\begin{equation}
	(\pd_1 + i \pd_2) S = 0.
\end{equation} 

If $m = 0$, there are two possible cases
\begin{equation}
	\begin{gathered}
		A_+ = 0,
		\qquad
		(\pd_0 + \pd_3) S = 0,
	\\
		B_+ = 0,
		\qquad
		(\pd_0 - \pd_3) S = 0.
	\end{gathered}
\end{equation} 

If $m \neq 0$, a new equation can be obtained by multiplying the first equation by $A_+$ and the third by $B_+$, and subtracting:
\begin{equation}
	2 A_+ B_+ \pd_0 S + m (A_+^2 + B_+)^2 = 0
\end{equation} 
such that
\begin{equation}
	m \, \left( \frac{A_+}{B_+} \right)^2 + 2 \pd_0 S \,  
	\frac{A_+}{B_+} + m = 0,
\end{equation} 
which admits for solution
\begin{equation}
	\frac{A_+}{B_+} = \frac{1}{m} \left( - \pd_0 S \pm \sqrt{(\pd_0 S)^2 - m^2} \right).
\end{equation}

\printbibliography[heading=bibintoc]

\end{document}